%% file: main.tex
\documentclass[sigconf, screen]{acmart}
\copyrightyear{2024}
\acmYear{2024}

\AtBeginDocument{%
  \providecommand\BibTeX{{%
    \normalfont B\kern-0.5em{\scshape i\kern-0.25em b}\kern-0.8em\TeX}}}

\usepackage[utf8]{inputenc}
\usepackage[T1]{fontenc}
\usepackage{microtype}
\usepackage[ruled, linesnumbered]{algorithm2e}
\usepackage{fancyhdr,graphicx,amsmath,eqnarray}
\usepackage{caption} 
\usepackage{xcolor}
\usepackage{graphicx}
\usepackage{balance}
\usepackage{subfigure}
\usepackage{multirow}
\usepackage{tikz}
\usepackage{pgfplots}
\usepackage{filecontents}
\usepackage{threeparttable}
\usepackage{booktabs}
\usepackage{enumitem}
\usepackage{makecell}
\usepackage{flushend}
\pgfplotsset{compat=1.13}
\usetikzlibrary{shapes, shapes.multipart, arrows.meta, positioning, matrix, calc, fit, decorations.pathreplacing, pgfplots.dateplot}
\usepgfplotslibrary{dateplot, statistics}

\newcommand{\ie}{{\em i.e.},\xspace}
\newcommand{\eg}{{\em e.g.},\xspace}
\newcommand{\MT}{Meituan\xspace}
\newcommand{\MIG}{{\textsf{AUITestAgent}}\xspace}

\begin{document}


\title{AUITestAgent: Automatic Requirements Oriented GUI Function Testing}


\author{Yongxiang Hu}
\authornote{This work was performed during the authors'  internship in Meituan.}
\affiliation{%
  \department{School of Computer Science}
  \institution{Fudan University}
  \city{Shanghai}
  \country{China}}

\author{Xuan Wang}
\authornotemark[1]
\affiliation{%
  \department{School of Computer Science}
  \institution{Fudan University}
  \city{Shanghai}
  \country{China}}

\author{Yingchuan Wang}
\authornotemark[1]
\affiliation{%
  \department{School of Computer Science}
  \institution{Fudan University}
  \city{Shanghai}
  \country{China}}

\author{Yu Zhang}
\affiliation{%
  \department{}
  \institution{Meituan}
  \city{Shanghai}
  \country{China}}

\author{Shiyu Guo}
\affiliation{%
  \institution{Meituan}
  \city{Shanghai}
  \country{China}}

\author{Chaoyi Chen}
\affiliation{%
  \institution{Meituan}
  \city{Beijing}
  \country{China}}

\author{Xin Wang}
\affiliation{
  \department{School of Computer Science}
  \institution{Fudan University \\
  Shanghai Key Laboratory of Intelligent Information Processing}
  \city{Shanghai}
  \country{China}}
  
\author{Yangfan Zhou}
\affiliation{
  \department{School of Computer Science}
  \institution{Fudan University \\
  Shanghai Key Laboratory of Intelligent Information Processing}
  \city{Shanghai}
  \country{China}}
  
\begin{abstract}

The Graphical User Interface (GUI) is how users interact with mobile apps.
To ensure it functions properly, testing engineers have to make sure it functions as intended, based on test requirements that are typically written in natural language.
While widely adopted manual testing and script-based methods are effective, they demand substantial effort due to the vast number of GUI pages and rapid iterations in modern mobile apps. 
This paper introduces \MIG, the first automatic, natural language-driven GUI testing tool for mobile apps, capable of fully automating the entire process of GUI interaction and function verification.
Since test requirements typically contain interaction commands and verification oracles.
\MIG can extract GUI interactions from test requirements via dynamically organized agents. Then, \MIG employs a multi-dimensional data extraction strategy to retrieve data relevant to the test requirements from the interaction trace and perform verification.
Experiments on customized benchmarks~\footnote{\href{https://github.com/bz-lab/AUITestAgent}{https://github.com/bz-lab/AUITestAgent}} demonstrate that \MIG outperforms existing tools in the quality of generated GUI interactions and achieved the accuracy of verifications of 94\%.
Moreover, field deployment in \MT has shown \MIG's practical usability, with it detecting 4 new functional bugs during 10 regression tests in two months.

\end{abstract}


\maketitle

\input{tex/1_Introduction}

\input{tex/2_Background}

\input{tex/3_Approach}
\input{tex/4_Evaluation}

\input{tex/5_Discussion}
\input{tex/6_Related}

\section{Conclusion}
\label{sec:conclusion}
In this paper, we propose \MIG, an automatic approach to perform natural language-driven GUI testing for mobile apps. 
In order to extract GUI interactions from test requirements, \MIG utilizes dynamically organized agents and constructs multi-source
input for them. 
Following this, a multi-dimensional data extraction strategy is employed to retrieve data relevant to the test requirements from the interaction trace to perform verifications.
Our experiments on customized benchmarks show that \MIG significantly outperforms existing methods in GUI interaction generation and could recall 90\% injected bugs with a 4.5\% FPr.
Furthermore, unseen bugs detected from \MT show the practical benefits of using \MIG to conduct GUI testing for complex commercial apps.
These findings highlight the potential of \MIG to automate the GUI testing procedure in practical mobile apps.




\newpage

\bibliographystyle{ACM-Reference-Format}
\bibliography{ui_testing}

\end{document}

%% file: tex/1_Introduction.tex
\section{Introduction}
\label{sec:intro}

Since GUI (Graphical User Interface) is the medium through which users interact with mobile apps, developers need to verify that it functions as expected.
The most direct approach to this verification is to check the UI functionality according to requirements, specifically by executing GUI interactions and checking GUI responses.
Despite being effective, the vast number of GUI pages and the rapid pace of app iterations result in a large volume of testing tasks. 
Therefore, to avoid expending a large amount of human effort, automation is essential.

Since the test requirements are usually expressed in natural language, we call this form of testing \textit{natural language-driven GUI testing} for mobile apps, which remains an underexplored field.
The most relevant work is natural language-driven GUI interaction\cite{DBLP:conf/icse/FengC24,DBLP:journals/corr/abs-2311-07562,DBLP:journals/corr/abs-2401-16158}. They typically utilize large language models (LLMs) to analyze natural language commands and GUI pages, and ultimately generate interaction commands.
However, unlike app interaction tasks, which focus on the interaction result and are goal-oriented, GUI testing requires checking the correctness of GUI responses during the interaction process and is, therefore, step-oriented.
Consequently, the commonly used trial-and-error strategies in these interaction methods often result in inefficient interaction traces and insufficient success rate for GUI testing.

In industry practices, performing \textit{natural language-driven GUI testing} still largely relies on intensive manual efforts. 
For example, such GUI testing in \MT is mainly performed manually or through GUI testing scripts. 
Specifically, testing engineers should read and understand test requirements and manually select UI elements to interact and check by XPath, a language for pointing to different parts from a UI hierarchy file ~\cite{url:xpath}. Then, multiple rules are designed to conduct GUI interactions and verifications.
However, mobile apps today typically have multiple business lines (\eg, hotel booking, taxi booking, and food delivery), each involving a tremendous number of GUI functions and rapid development iterations. Consequently, conducting GUI function testing demands substantial manual effort for test script development and maintenance.
This paper aims to reduce such manual efforts in \textit{natural language-driven GUI testing}. 

Despite GUI testing having more steps and therefore being more complex than interactions for testing engineers, we found that the step-oriented characteristic of GUI testing tasks is advantageous.
This structure allows us to utilize step descriptions to design prompts, enabling us to achieve higher UI interaction success rates and verification accuracy, thereby attaining industry-level usability.

Although encouraging, there are still challenges posed by diverse testing requirements and the complex, huge amount of GUI pages. The key to reducing such human efforts lies in simplifying GUI testing tasks to align with LLMs' capabilities.
In this regard, we designed \MIG, the first automatic natural language-driven GUI testing tool for mobile apps. It takes GUI testing requirements as input, performs tests on the specified app, and outputs the test results.

As for \MIG's implementation, three strategies are employed to simplify GUI testing tasks. 
Firstly, although GUI testing contains GUI interaction and function verification, the verification results do not actually impact the interaction. Therefore, \MIG decouples interaction and verification into two separate modules, performing verification after interaction.
As for diverse interaction commands, we designed two agent organization patterns. For simple instructions, \MIG allows the \textit{Executor} to perform directly, while complex instructions are broken down by the \textit{Planner} first.
As for GUI function verification, it requires processing numerous GUI screenshots, each containing tens of GUI elements. Since LLMs struggle to handle questions with many images attached accurately\cite{DBLP:conf/aaai/HuXLLCT24,DBLP:journals/tacl/LiuLHPBPL24}, \MIG extracts requirement-related information from three dimensions: GUI elements, GUI pages, and interaction traces. 
Then, verification will be conducted based on this integrated information.

We analyze the design effectiveness of \MIG with a set of experiments. We prove that \MIG perform the best in GUI interaction compared with AppAgent~\cite{DBLP:journals/corr/abs-2312-13771} and MobileAgent~\cite{DBLP:journals/corr/abs-2401-16158}, which are both well-known natural language-driven interaction tools.
Our verification experiment shows that \MIG can recall 90\% GUI functional bugs and provide reasonable explanations while maintaining a false positive rate of less than 5\%.
We also report our field experiences in applying \MIG in \MT. To date, \MIG has been deployed across xx business lines and has detected xx new functional bugs.
We summarize the contributions of this paper as follows.
\begin{itemize}[leftmargin=*]
    \item We demonstrate that the step-oriented characteristic of GUI functional testing is well-suited for LLMs-based automatic testing.
    \item We propose \MIG, the first automatic natural language-driven GUI testing tool for mobile apps.
    \item We show that \MIG is an effective tool via real-world cases on commercial apps that serve tremendous end-users. We summarize the lessons learned, which can shed light on the automation of GUI testing.
\end{itemize}

%% file: tex/2_Background.tex
\section{Practices of GUI Function Testing}
\label{sec:background}

\input{figure/2_test_requirement}
The design of \MIG is motivated by the practical testing status of \MT's mobile app. \MT is one of the largest online shopping platforms over the world, with nearly 700 million transacting users and about 10 million active merchants. Bugs on GUI pages in such apps would inevitably deteriorate user experience.

The \MT app features a vast number of GUI pages and functionalities. 
Despite numerous automated testing tools proposed in academia~\cite{url:monkey,DBLP:conf/icse/GuSMC0YZLS19,DBLP:conf/sigsoft/SuMCWYYPLS17,DBLP:conf/issta/MaoHJ16,DBLP:conf/sigsoft/HuZY18,DBLP:conf/icse/CaoP000WL24}, they typically focus on detecting non-functional bugs (\eg app crash) or require considerable manual intervention.
Consequently, these tools fail to reduce the human effort required for GUI functional testing substantially.

During our practice in \MT, GUI function verification is mainly performed manually or through hard-coded scripts. 
Specifically, testing engineers need to collaborate with both requirements and development teams to define test requirements.
As shown in Figure~\ref{fig:req}, these requirements are typically in natural language.
After the GUI development is completed, testing engineers verify these test requirements manually on real devices. 
Additionally, regression tests are set for core functionalities for cost considerations. 
Implementing GUI function regression testing involves developing hard-coded GUI testing scripts. This requires testing engineers to select UI elements that need interaction or verification from the UI hierarchy~\cite{url:UI_layout} and perform GUI interaction and function verification via rule development.

\input{figure/3_overview}

For commonly used commercial apps, there are usually many business lines, each with numerous functions, and each function involves multiple GUI interfaces. 
As a result, the workload for GUI functional testing is substantial. 
Although GUI test scripts can reduce some of the manual effort, they are prone to becoming outdated due to GUI changes brought by app iterations. Therefore, extensive use of these scripts can lead to considerable maintenance costs.

%% file: figure/2_test_requirement.tex
\begin{figure}[t!]
    \centering
    \subfigure[Requirement with specific steps]{
        \begin{minipage}[t]{0.47\linewidth}
        \centering
        \includegraphics[width=\linewidth]{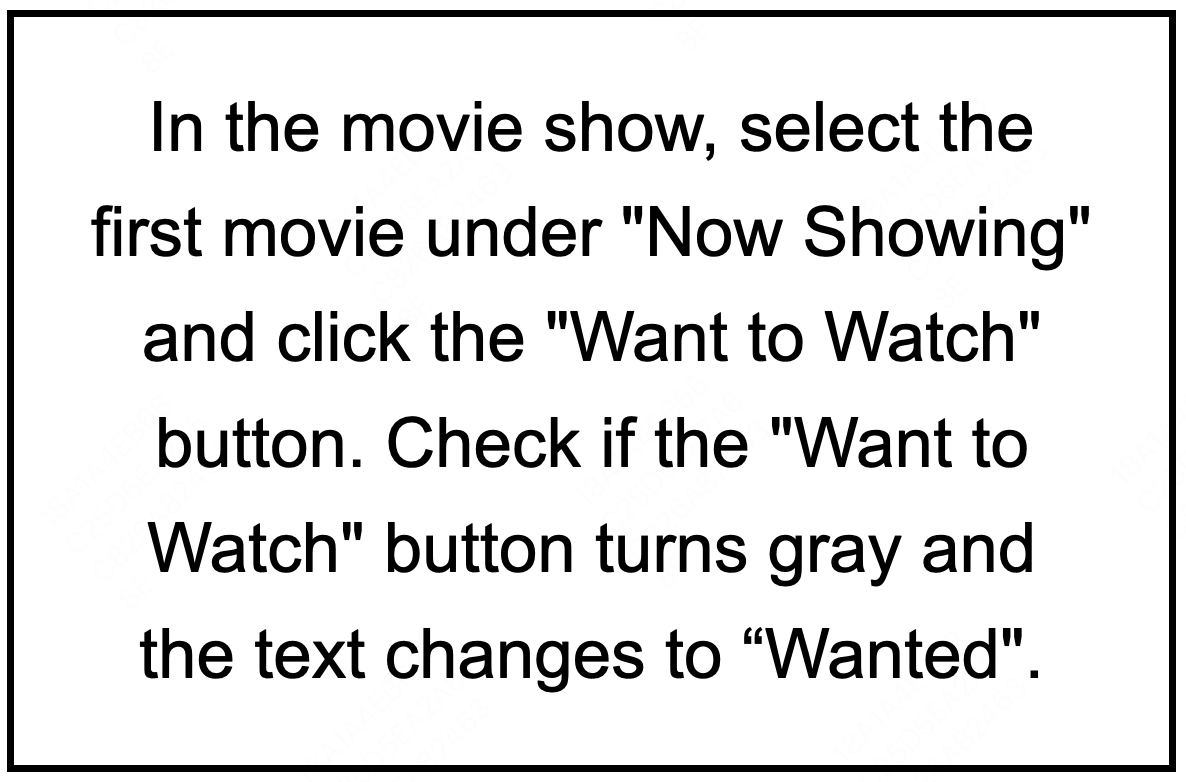}
        \label{fig:req_1}
        \end{minipage}
    }
    \subfigure[Requirement in concise expression]{
        \begin{minipage}[t]{0.47\linewidth}
        \centering
        \includegraphics[width=\linewidth]{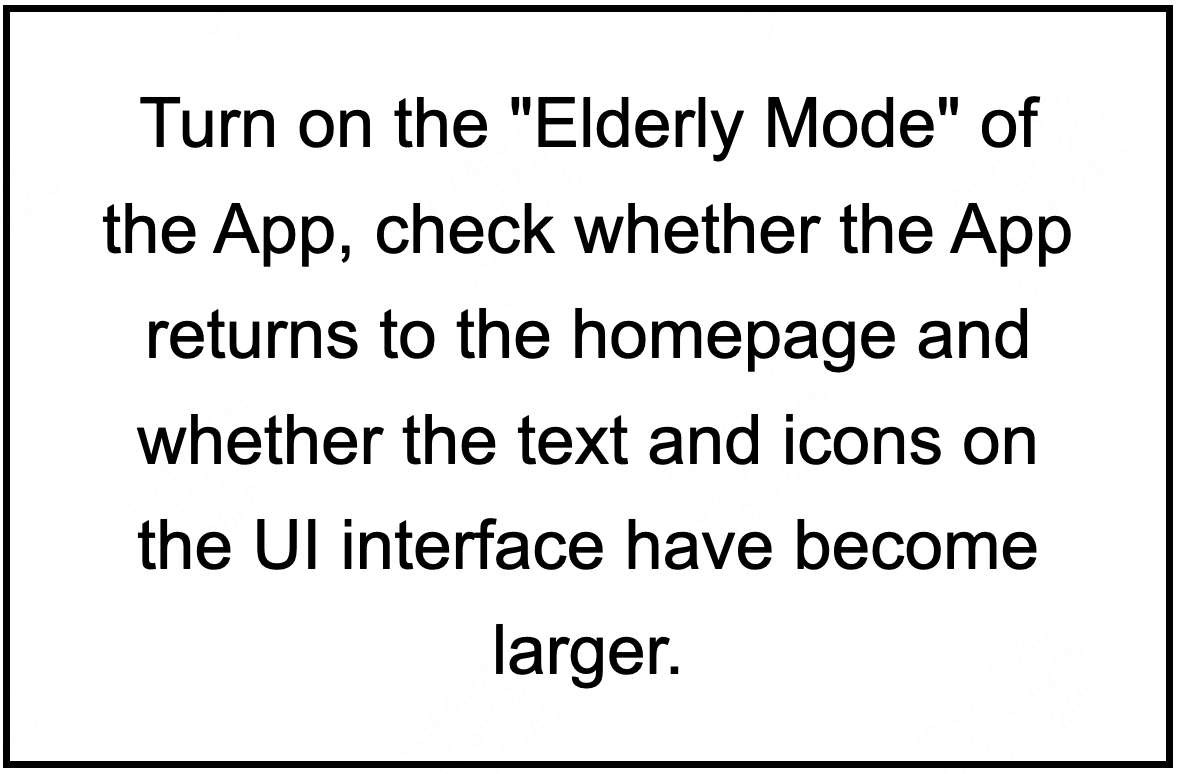}
        \label{fig:req_2}
        \end{minipage}
    }
    \caption{Practical Testing Requirements}
    \label{fig:req}
\end{figure}

%% file: figure/3_overview.tex
\begin{figure*}
    \centering
    \includegraphics[width=0.8\textwidth]{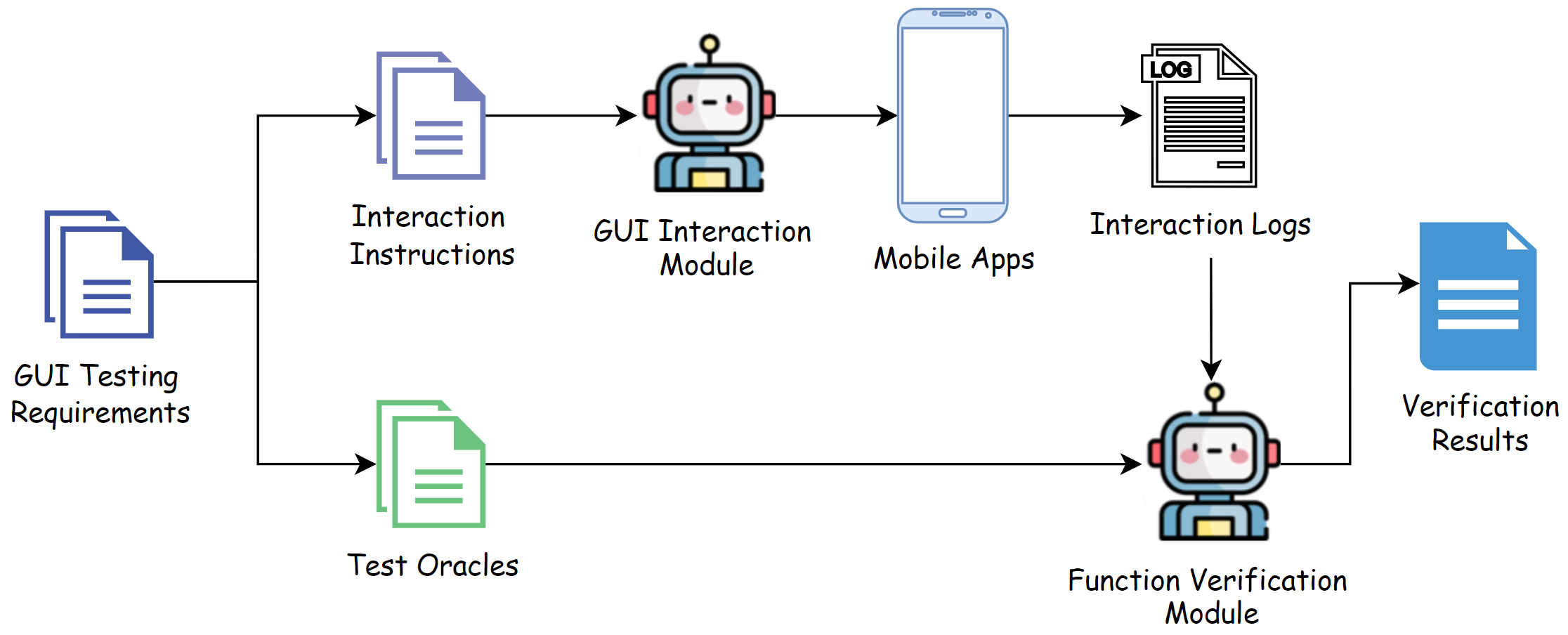}
    \caption{Overview of \MIG}
    \label{fig:overview}
\end{figure*}

%% file: tex/3_Approach.tex
\section{Tool Design and Testing Workflow}
\label{sec:method}

\subsection{Overview}
\label{ssec:overview}

Figure~\ref{fig:overview} presents an overview of \MIG, which comprises two main modules: GUI interaction and function verification. 
Specifically, \MIG first analyzes the GUI test requirements to extract interaction commands and oracles for function verification.
The GUI interaction module then dynamically selects the appropriate agent organization pattern based on the complexity of the interaction commands. 
If the commands consist of specific steps, the interactions will be performed directly by the \textit{Executor}. Otherwise, if the commands are expressed concisely, a more complex pattern involving a \textit{Planner} and \textit{Monitor} will be used.
Finally, the function verification module analyzes the interaction logs based on the test oracles. These logs contain information about the GUI pages encountered during the interactions, including screenshots, LLM outputs, and the UI actions performed. 
\MIG extracts oracle-relevant data from the interaction logs and outputs the verification results and reason analysis.

\subsection{Natural Language Driven GUI Interaction}
\label{ssec:interaction}
\input{figure/3_interaction}

As discussed in Section~\ref{sec:background}, the interaction commands in different GUI test requirements vary significantly. 
As shown in Figure~\ref{fig:req_1}, target elements and corresponding UI actions are specified in explicit commands.
In contrast, concise commands such as Figure~\ref{fig:req_2} may abstract multiple UI actions into a single interaction command, which is more challenging to perform.
Such differences require \MIG to dynamically organize suitable agents to collaborate, ensuring the interaction commands are executed accurately.

Specifically, concise interaction commands require the agent to accurately interpret the user's intent behind the commands, assess the current state from GUI screenshots and interaction records, and determine the appropriate UI actions to take. 
Therefore, performing concise commands demands interaction memory and reasoning capabilities, which are difficult to achieve through a single agent due to the context length limitations.
In contrast, while detailed interaction commands are easier to execute, using multiple agents increases unnecessary costs. It may also amplify the hallucination problem~\cite{DBLP:journals/csur/JiLFYSXIBMF23} of LLMs due to cumulative effects, thereby reducing the quality of generated UI actions.

To address these challenges, the GUI interaction module of \MIG employs two agent organization patterns,\textit{Specific steps pattern} and \textit{Concise expression pattern}, as shown in Figure~\ref{fig:interaction}. 
The GUI interaction module first uses LLMs to recognize the type of input interaction commands. If the interaction commands consist of a series of specific steps, \MIG will convert them into a list of single-step UI actions via the specific step pattern. 
Otherwise, if the test requirement contains concise interaction commands, the GUI interaction module will use the concise expression pattern to handle them.

\subsubsection{Specific steps pattern}
\label{sssec:specific_interaction}
Specific steps pattern executes each operation in the operation list sequentially through the collaboration of three LLM agents: the \textit{Observer}, the \textit{Selector}, and the \textit{Executor}.

The \textbf{Observer} agent is to identify all interactive UI elements in the current GUI page and infer their functions. 
Specifically, the \textit{Observer} processes the screenshot of the current GUI page using Set-of-Mark\cite{DBLP:journals/corr/abs-2310-11441}, assigning numeric IDs and rectangular boxes to all interactive elements.
This approach highlights the interactive elements in the GUI screenshots, making it easier for LLMs to recognize.
Finally, a multi-modal large language model (MLLM) is used to infer the function of each numbered UI element.

Although MLLMs possess image analysis capabilities, they do not perform well in observing GUI images, especially in recognizing UI elements and analyzing their functions due to the complexity of GUIs\cite{DBLP:conf/sigsoft/HuGHZTGCZ23}. 
Besides, MLLMs are prone to Optical Character Recognition (OCR) hallucinations when recognizing text in images\cite{DBLP:journals/corr/abs-2402-14683}, while text is crucial for understanding the functionality of many elements within the GUI.
To tackle these challenges, we enhance the GUI analysis capability of the \textit{Observer} agent through multi-source input and knowledge base augmented generation.

Specifically, we select GUI screenshots and UI hierarchy file~\cite{url:UI_layout} as the \textit{Observer}'s input.
For the UI hierarchy file, \textit{Observer} first parses the UI hierarchy file to identify all the interactive nodes(\ie whose clickable, enabled, scrollable, or long-clickable attribute is set to true). 
Related attributes of these nodes are also collected, including coordinate information (from `bounds' attribute), textual information (from `text' and `content-desc' attribute), and functional information (\eg nodes with a ``EditText'' class are likely to be text input fields). 
For GUI screenshots, \textit{Observer} uses the vision-ui model from \MT \cite{url:vision-ui} to identify all elements in the GUI and perform OCR to recognize the text visible to users.
The recognition results serve as a supplement to the UI hierarchy, especially in cases where obtaining the UI hierarchy file fails or when the page contains WebView~\cite{url:webview} components. 

In order to make it easier for LLMs to describe the functions of elements, the \textit{Observer} then marks the numerical ID and rectangular boxes of each element on the GUI image based on their coordinates. 
Finally, a system prompt that contains the processed screenshot and corresponding elements' data, including element IDs, text information, and other property information, is assembled. 
This prompt is then used as input to the MLLM to infer the function of UI elements.

Given that the LLM is not well-acquainted with the design of mobile apps, particularly those in Chinese, we have observed that the \textit{Observer} agent occasionally mispredicts the functions of UI elements, especially those UI icons that lack textual information.
To tackle this problem, a UI element function knowledge base is formed to enhance element recognition.
For instance, we manually collect and label some elements that are difficult to recognize, primarily icons that do not contain text.
Each element in the knowledge base contains its screenshot, appearance descriptions, and function. 
When processing the GUI, the \textit{Observer} uses CLIP~\cite{DBLP:conf/icml/RadfordKHRGASAM21} to match elements from the knowledge base with elements on the current page and then includes the matched elements' appearance descriptions and functions in the system prompt for MLLM. This helps it infer the functions of these error-prone elements.

The \textbf{Selector} agent is set to select the operation target from the elements listed by \textit{Observer} based on the natural language description of a single-step command. 
It generates the necessary parameters for the GUI action and then calls the corresponding function to perform. Specifically, we have defined the following UI actions for \textit{Selector}:
\begin{itemize}[leftmargin=*]
    \item click(target: int): This function is used to click an element listed by \textit{Observer}. The parameter target is the element's ID.
    \item longPress(target: int): This function is similar to click, while its function is to long press a particular element.
    \item type(target: int, text: str): this function is used to type the string `text' into an input field. It is worth noting that the target parameter is the ID of the input field. The function will click on this element before typing to ensure the input field is activated.
    \item scroll(target: int, dir: str, dist: str): This function is called to scroll on the element target in a direction `dir'(up, down, left, right) and distance `dist'(short, medium, long). 
\end{itemize}
Finally, the \textbf{Executor} agent performs the UI actions generated by \textit{Selector} on the app under testing (AUT). 
For actions targeting specific elements, the \textit{Executor} first consults the coordinates provided in the \textit{Observer}'s results, converting the ID of the target element into specific page coordinates.
Then \textit{Executor} performs the actual interaction action through the Android Debug Bridge (adb) tool\cite{url:adb}. 

\input{figure/3_verification}

\subsubsection{Concise expression pattern}
Concise expression pattern is also achieved through the collaboration of multiple agents. As shown in Figure~\ref{fig:interaction}, it adds two agents, \textit{Planner} and \textit{Monitor}, on top of the specific step pattern to handle concise commands. 

The \textbf{Planner} agent formulates a plan based on the concise interaction commands, where each step in the plan is a specific interaction command aimed at progressing the completion of the interaction commands.
Then, for each step in the plan, the specific step pattern is used for GUI action execution.

Another agent, the \textbf{Monitor}, is involved because, unlike the specific step pattern that requires only the sequential execution of each operation in the list, the concise expression pattern necessitates a role to determine whether the user's commands have been completed (\ie whether the user's intent has been achieved).

Specifically, the \textit{Monitor} determines whether the interaction commands have been completed based on the current GUI screenshot and the record of executed operations. If the commands are completed, the GUI interaction module outputs the interaction log and stops. 
Otherwise, the \textit{Monitor} describes the current GUI state and the operations it believes are necessary to complete the rest interactions. Its output serves as feedback for the \textit{Planner}, and in the next iteration, the \textit{Planner} will make a new plan.

\subsection{Function Verification}
To our knowledge, \MIG is the first work to investigate GUI function verification. 
Since \MIG decouples GUI interaction from verification, its verification is conducted based on interaction log analysis.
Consequently, we formulate function verification as a semantic analysis problem of GUI pages.

Although MLLMs possess image analysis capabilities, we still need to simplify GUI function verification tasks to align with the capabilities of LLMs, to achieve stable performance.
Specifically, there are three challenges in GUI function verification. First, the diversity of test oracles makes them difficult to process.
Second, the sheer volume of data in interaction logs presents a challenge, as LLMs struggle to process excessively lengthy inputs~\cite{DBLP:conf/aaai/HuXLLCT24}, which may even exceed their maximum input capacity.
Lastly, according to our experience, even the most capable MLLMs are not adept at semantic analysis of multiple images, especially complex images like GUI screenshots.

\subsubsection{Simplifying Test Oracles}
To tackle challenges posed by test oracles' complexity, \MIG will convert them to a list of simple questions, termed \textit{verification points} in this paper.
According to our experiences in \MT, test requirements often encompass multiple verification points, with each relating to different pages and UI components within the interaction log.
As shown in Figure~\ref{fig:verify}, \MIG initially breaks down the test oracles extracted from the test requirements into individual verification points.
Each verification point is tailored to address a single question, thereby streamlining the handling of test oracles.

\subsubsection{Multi-dimensional Data Extraction}

As for the challenges posed by the scale of log data and semantic analysis of images, \MIG approaches information extraction from three perspectives, transforming it into textual information. 
Specifically, data contained in interaction logs can be categorized into three dimensions, as shown in Table~\ref{table:dimensions}.
The GUI element dimension contains element functions and properties such as location and shape.
Another dimension, the GUI page, includes page types (\eg form page and list page) and categories (\eg payment page, advertising page).
And the page state transitions dimension, integrating the first two aspects (\eg a payment page can be reached from an order-placement page by clicking \textit{Confirm} button).

\input{table/3_multi-demensional}

\MIG leverages multiple agents to extract information relevant to the current verification point from these dimensions and records it as text. 
Considering the analysis of GUI element functions and page types has been performed in Section~\ref{ssec:interaction}, \MIG will reuse such content to enhance processing speed.

\subsubsection{Function Verification}
After requirement-related data are collected, \MIG will integrate the verification points with corresponding data extracted from the logs. 
Finally, the judgment will be executed using MLLM.
Furthermore, we employ the JSON mode~\cite{url:JSONmode} to standardize the output, ensuring a structured delivery of judgments and explanations.

%% file: figure/3_interaction.tex
\begin{figure*}
    \centering
    \includegraphics[width=\textwidth]{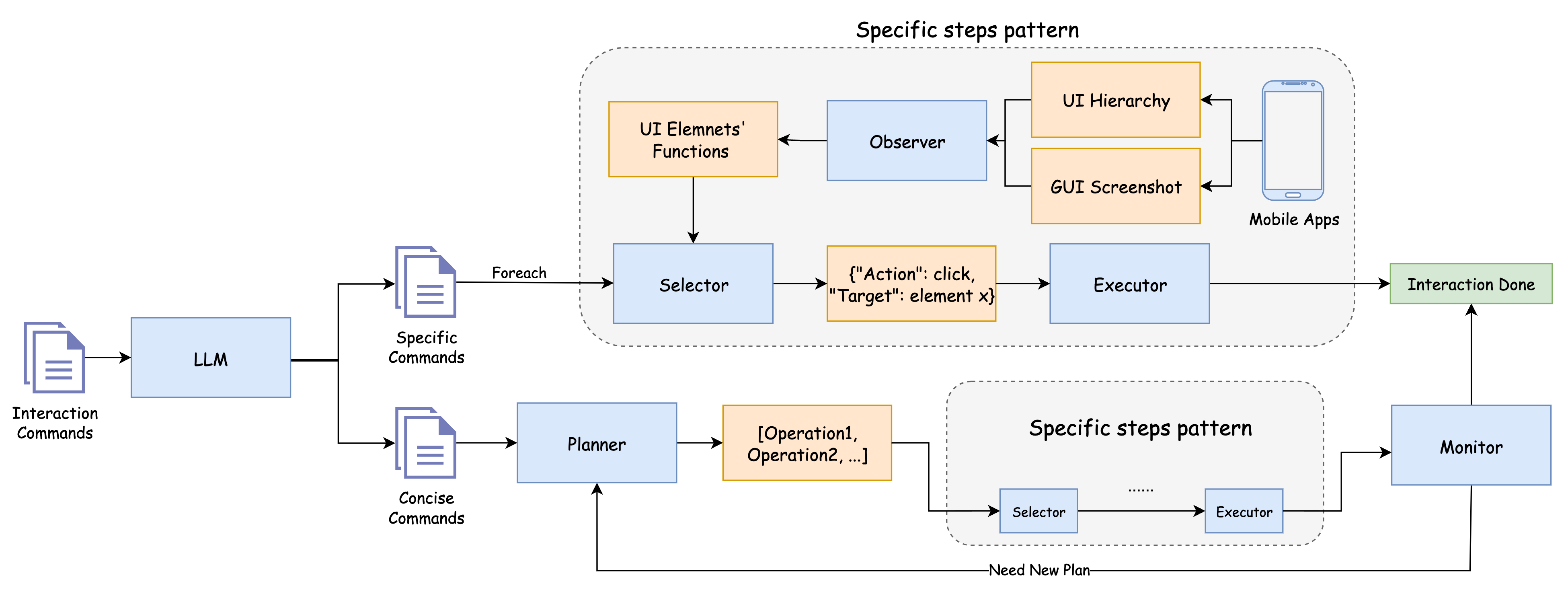}
    \caption{Workflow of GUI Interaction Module}
    \label{fig:interaction}
\end{figure*}

%% file: figure/3_verification.tex
\begin{figure*}
    \centering
    \includegraphics[width=0.9\textwidth]{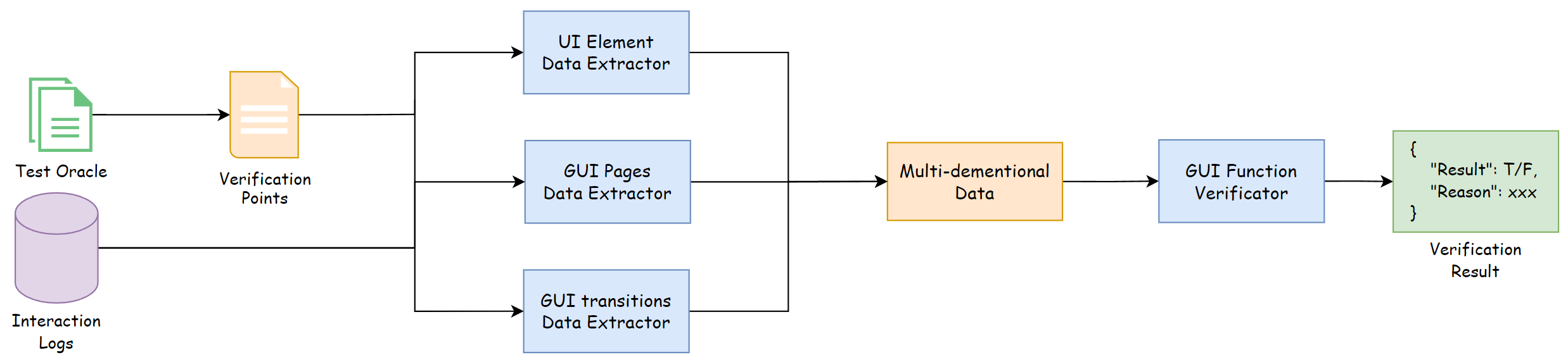}
    \caption{Workflow of Function Verification Module}
    \label{fig:verify}
\end{figure*}

%% file: table/3_multi-demensional.tex
\begin{table}[t!]
\centering
\caption{Categorization of interaction log data.}
\vspace{-6pt}
\label{table:dimensions}
\resizebox{\linewidth}{!}{
    \begin{threeparttable}
    \begin{tabular}{c|c|c}
    \hline
    \textbf{Type} & \textbf{Description} & \textbf{Example} \\
    \toprule
    UI element & Element functions and properties & A submit button at the bottom-right corner. \\
    \midrule
    GUI page & Page structures and categories & A list-structured advertisement page. \\
    \midrule
    Page state & \multirow{2}{*}{integrate the first two aspects.} & A payment page can be reached from an order- \\
    transitions & & placement page by clicking the submit button.\\
    \bottomrule
    \end{tabular}
    \end{threeparttable}
}
\end{table}

%% file: tex/4_Evaluation.tex
\section{Evaluation}
\label{sec:evaluation}

In this section, we evaluate the performance of \MIG by answering the following research questions.

\begin{itemize}[leftmargin=*]

\item \textbf{RQ1:} How do the quality and efficiency of \MIG in generating GUI interactions compare to the existing approaches?
\item \textbf{RQ2:} How do the effectivness of \MIG in perform GUI function verification?
\item \textbf{RQ3:} To what extent can \MIG perform natural language-driven testing in practical commercial mobile apps?

\end{itemize}

RQ1 assesses the effectiveness of \MIG's GUI interaction model, examining its accuracy and efficiency in converting natural language commands into GUI actions. 
In RQ2, we construct a dataset containing 20 bugs, simulating real-world scenarios to evaluate the effectiveness of the GUI function verification module. 
In RQ3, we validate the practical applicability of \MIG by examining its performance across GUI pages in \MT.

\subsection{Evaluation Setup}
\subsubsection{Dataset}

\input{table/4.1_interaction_tasks}
\input{table/4.1_Oracles}
To the best of our knowledge, \MIG is the first automatic natural language-driven GUI testing tool for mobile apps. Since there is no publicly available dataset, we contract testing scenarios and corresponding requirements for RQ1-RQ2. 
As shown in Table~\ref{table:interaction-tasks}, given the diversity of practical test requirements depicted in Figure~\ref{fig:req}, we construct interaction requirements of various complexities along with corresponding verification oracles of various types.

For RQ1, we assessed the difficulty of interactions from two perspectives: the ideal number of interaction steps (\ie the number of steps required for manual interaction, recorded as $Step_{ideal}$) and the vagueness of the requirements (\ie metric $Score_{vag}$ is defined as the $Step_{ideal}$ divided by the number of interaction commands in the requirements).
As shown in Table~\ref{table:interaction-tasks}, we use $Step_{ideal}$ and $Score_{con}$ to describe the complexity level of interaction tasks.
Based on these two perspectives, we have categorized the interaction tasks into three difficulty levels. We then created ten interaction tasks for each difficulty level from eight popular apps \ie \MT, Little Red Book, Douban, Facebook, Gmail, LinkedIn, Google Play and YouTube Music.

For RQ2, as the apps we selected are not open-source and bugs are not common in the online versions of these mature commercial apps, we needed a robust method to effectively measure \MIG's performance in executing function verification.
Two authors of this paper analyzed 50 recent historical anomalies at \MT, categorizing them into three types (as shown in Table~\ref{table:verification-points}). Based on these categories, we manually constructed 20 test oracles. 
Specifically, we selected 20 interaction tasks from the 30 crafted for RQ1 and manually created corresponding GUI anomaly manifestations. Consequently, as shown in Table~\ref{table:verification-points}, the data for RQ2 includes 20 test oracles, with each oracle associated with two interaction traces, one correct and the other having conflict with a verification point. 
Therefore, there are 40 verification tasks for RQ2, half of which contain anomalies.

For RQ3, we applied \MIG to 18 test requirements from the real business scenarios at \MT.
By presenting the bugs identified by \MIG in real scenarios, we demonstrate its practical effectiveness.

\subsubsection{Baselines}
\input{table/4.2_Interaction_perf}
\input{figure/4_rq2baseline}
To demonstrate the advantages of \MIG in converting natural language into GUI interaction, we compared it with two state-of-the-art natural language-driven GUI interaction tools. 
AppAgent~\cite{DBLP:journals/corr/abs-2312-13771} is the first to propose a natural language-driven approach for mobile app interaction, featuring an innovative design that incorporates an exploratory learning phase followed by interaction. 
This addresses the challenge of LLMs' lack of familiarity with apps by offering two learning modes: unsupervised exploration and human demonstration. 
For fairness, in RQ1, we utilized the unsupervised exploration mode and set the maximum number of exploration steps to $Step_{ideal}+5$.
Different from AppAgent, MobileAgent~\cite{DBLP:journals/corr/abs-2401-16158} employs a visual perception module for operation localization based solely on screenshots without needing underlying files.
It utilizes LLM agents for holistic task planning and incorporates a self-reflection method to correct errors and halt once tasks are completed.

Since \MIG is the first to focus on natural language-driven GUI function verification and there are no existing studies in this field, we constructed a verification method based on multi-turn dialogue~\cite{DBLP:conf/aaai/XuZ021} as a baseline.
Specifically, We constructed a prompt of a two-turn dialogue with MLLM, as shown in figure \ref{fig:rq2_baseline}.

\subsubsection{Implementation and Configuration}
For RQ1 and RQ2, to explore the upper limits of \MIG's capabilities, we selected one of the state-of-the-art MLLM, GPT-4o~\cite{url:gpt4-o}, as the underlying model for \MIG and three baselines.
As for \MIG's practical implementation within \MT, we select another MLLM.
Optimizing for efficiency, and given that all of these methods are device-independent, we conducted our experiments on multiple Android devices via the adb tool. 
For fairness, in RQ1, we set the maximum number of trial steps for all candidates to $Step_{ideal}+5$.

\subsection{RQ1: Quality of GUI Interactions}
\input{table/4.3_veri_perf}
Initially, three authors manually executed each of the 30 interaction tasks. 
For each task, we recorded the manual interaction trace as $\text{Trace}^{\text{ma}}_i$, which serves as the ground truth for interaction task $i$. 
\begin{equation}
    \text{Trace}^{\text{ma}}_i = \{\text{action}_1, \dots, \text{action}_k\}
\end{equation}
The ideal number of interaction steps, $\text{Step}_{\text{ideal}}$, is defined as the length of $\text{Trace}^{\text{ma}}_i$.
For the action sequence generated by tool $x$, we denote it as $\text{Trace}^x_i$.
\begin{equation}
    \text{Trace}^x_i = \{\text{action}_1, \dots, \text{action}_m\}
\end{equation}
Due to the trial-and-error interaction strategy of the baseline interaction tool, we incorporated a parser in RQ1 to filter out redundant actions (\ie revert and repeat the last action) and meaningless interactions (\eg do not click on a UI element) in $\text{Trace}^x_i$. 
The filtered result is denoted as $\text{Trace\_P}^x_i$.

Four metrics are used to evaluate the interaction quality of \MIG and baselines.
\begin{itemize}[leftmargin=*]
    \item \textit{Task Completion}: Abbreviated as TC, indicates whether interaction tasks have been successfully executed.
    \item \textit{Correct Step} (CS): Represents the correct actions in $\text{Trace}^x_i$.
    \item \textit{Correct Trace} (CT): Represents the longest common prefix between $\text{Trace}^x_i$ and the ground truth \ie $\text{Trace}^{\text{ma}}_i$. It can indicate the degree of completion of the interaction task $i$.
    \item \textit{Step Efficiency} (SE) is used to measure the interaction efficiency of the action traces generated by \MIG and baselines when task completion occurs. It is represented by the ratio of $\text{Trace\_P}^x_i$ to $\text{Trace}^x_i$.
\end{itemize}

The formal representations of these metrics are as follows:
\begin{align}
    TC_i &= \begin{cases} 
        1 & \text{if}\: \: \text{Trace\_P}^x_i == \text{Trace}^{\text{ma}}_i \\
        0 & \text{Otherwise}
    \end{cases}  \\[8pt]
    CS_i &= \frac{|\text{Trace}^x_i \cap \text{Trace}^{ma}_i|} {|\text{Trace}^{x}_i|} \\[8pt]
    CT_i &= \frac{|LCP(\text{Trace}^x_i, \text{Trace}^{ma}_i)|}{|\text{Trace}^{ma}_i|} \\[8pt]
    SE_i &= \begin{cases}
         |\text{Trace\_P}^x_i|\: /\: |\text{Trace}^x_i| & \text{if}\: \: TC_i == 1 \\
         0 & \text{Otherwise}
    \end{cases}
\end{align}

As shown in Table~\ref{table:interaction-perf}, \MIG achieved the highest scores across four types of metrics, with an overall task completion rate of 77\%, significantly outperforming the two baseline methods. 
Although AppAgent incorporates an exploratory process, its direct use of the unprocessed UI hierarchy as input for LLMs results in excessively long and noisy data due to the complex GUI interfaces of the popular apps selected for RQ1. 
Consequently, this approach leads to the lowest performance metrics for AppAgent across all evaluated criteria.
This also demonstrates the effectiveness of the multi-source GUI understanding implemented in \MIG.

Furthermore, as the difficulty of interaction tasks increases, four metrics for \MIG and the baselines decline. 
Although Mobile-Agent performs well at levels L1 and L2, it struggles with tasks that involve numerous interaction commands or require concise execution. 
This reflects the significance of the design in \MIG’s GUI interaction module.

Therefore, the conclusion of RQ1 is that \MIG can effectively convert GUI interaction tasks of different difficulty levels into corresponding GUI actions.
Since the following function verification requires that interaction tasks be completed first, \MIG possesses the capability to carry out subsequent function verification.

\subsection{RQ2: Effectiveness of Function Verification}

Since some test oracles in \MT contain multiple verification points, 8 of the 20 test oracles we designed for RQ2 include multiple verification points. 
Before conducting the experiments, we manually divided these test oracles into 32 verification points to facilitate subsequent analysis.

We evaluated the verification performance of \MIG and the multi-turn dialogue-based baseline from three perspectives: test oracle, verification point, and reasoning in explanation. To ensure fairness, we also tracked the number of tokens consumed by each method in the experiment. Specifically, we designed five metrics for RQ2.

\textit{Oracle Acc.} represents the proportion of correctly judged tasks. A task is considered correctly judged by an oracle only if all included verification points are accurately assessed and the corresponding explanations are reasonable. Similarly, \textit{Point Acc.} and \textit{Reasoning Acc.} measure the accuracy of verification point judgments and the correctness of their explanations.
Additionally, we present the \textit{Completion Tokens} and \textit{Prompt Tokens} for each task, which can indicate the economic costs associated with each method.

As shown in Table~\ref{table:veri-perf}, \MIG achieved an accuracy of 0.90 in making correct judgments on test oracles across two types of tasks.
Due to the greater difficulty of anomaly detection compared to correct function verification, \MIG achieved an \textit{Oracle Acc.} of 0.85 in anomaly detection, slightly lower than in correct function verification. 
Considering that the 40 cases in RQ2 contain 64 verification points, of which 20 have anomalies, \MIG successfully identified 18 anomalous points, resulting in 2 false positives. This corresponds to a recall rate of 90\% and a false positive rate of 4.5\%.

Conversely, the baseline method reached an accuracy of 1.0 in the correct function verification task but only achieved an \textit{Oracle Acc.} of 0.35 in anomaly detection, indicating its difficulty in understanding questions in RQ3 and its tendency to output a `no anomaly' judgment. 

This data not only highlights the challenges LLMs face in handling GUI functional testing but also validates the effectiveness of multi-dimensional data extraction in \MIG.

\subsection{RQ3: Performance in Practical Implementations}
\input{table/4.4_RQ3_Bug}

In RQ3, we discuss the practical implementation of \MIG in real business scenario \ie \textit{Video} of \MT's app, to show its practical effectiveness. 
Specifically, 10 test requirements with L1/L2 level interaction commands are selected to cover various checkpoints in short video playback and live show scenarios. 
These requirements have been integrated into the weekly regression automation process for this business line in \MT.

In the video scenario within \MT, in addition to supporting normal video playback functions (pause and play, dragging the progress bar, and swiping up and down to switch videos), it also supports clicking on merchant and product cards to view merchant information and purchase products. The GUI page in this scenario is complex, with numerous checkpoints that need testing. In the past, each iteration relied on manually writing test scripts or performing manual interactions for verification.
These 10 test requirements contain 18 verification points in the video scenario, saving 1.5 Person-Days (PD) in each round of regression testing. 
Since \MIG's launch, it has detected 4 new bugs from the video scenario of \MT's app during 10 rounds of regression testing. Details of these bugs are shown in Table ~\ref{table:rq3-bug}.

%% file: table/4.1_interaction_tasks.tex
\begin{table*}[h!]
\centering
\footnotesize
\renewcommand{\arraystretch}{1.2}
\caption{Difficulty levels of interaction tasks.}
\label{table:interaction-tasks}
\resizebox{1.0\linewidth}{!}{
\begin{tabular}{c|>{\centering\arraybackslash}m{4cm}|>{\centering\arraybackslash}m{5cm}|c|c}
  \hline
  \textbf{Difficulty Levels} & \textbf{Definition} & \textbf{Examples} & \textbf{$\text{Step}_{ideal}$} & \textbf{$\text{Score}_{vag}$} \\
  \hline
  \multirow{2}{*}[-4ex]{L1} & \multirow{2}{*}[-4ex]{\parbox{4cm}{\centering $ \text{Step}_{ideal}+\text{Score}_{vag} \leq 3$}} & Click the Like button on the first post. & 1 & 1 \\
  \cline{3-5}
  & & Click the Top Charts tab, then click to view the details of "Honor of Kings". & 2 & 1 \\
  \cline{3-5}
  & & Click the "Explore" tab, then click the "New releases". & 2 & 1 \\
  \hline
  \multirow{2}{*}[-4ex]{L2} & \multirow{2}{*}[-4ex]{\parbox{4cm}{\centering $3 < \text{Step}_{ideal}+\text{Score}_{vag} \leq 7$ }} & Click the "Movie Showings" button, click on the first movie under "Now Playing," and then click the "Want to Watch" button. & 3 & 1 \\
  \cline{3-5}
  & & Click the "Takeout" button, click on the "Food" section, and then click to enter the first store in the store list. & 3 & 1 \\
   \cline{3-5}
  & & View the detail page of the first notification in the notifications list. & 3 & 3 \\
  \hline
  \multirow{2}{*}[-8ex]{L3} & \multirow{2}{*}[-8ex]{\parbox{4cm}{\centering $ \text{Step}_{ideal}+\text{Score}_{vag} > 7$}} & Search for "Hello World" and then play any song from the search list. & 4 & 4 \\
  \cline{3-5}
  & & Click the "Jobs" tab, search for QA Engineer in the search bar, click to view the details of any job from the search results, and then click Save. & 6 & 1.5 \\
  \cline{3-5}
  & & Send an email to \{address1\} and \{address2\} with the subject "Test" and the body "Hello". & 8 & 2.66 \\
  \hline
\end{tabular}
}
\end{table*}

%% file: table/4.1_Oracles.tex
\begin{table*}[h!]
\centering
\small
\caption{The categories of test oracles.}
\vspace{-6pt}
\label{table:verification-points}

\begin{tabular}{c|c|c}
  \hline
  \textbf{Category} & \textbf{Definition} & \textbf{Examples}\\
  \hline
  \multirow{2}{*}{No Response} & \multirow{2}{*}{App does not react when it should have.} & No change after clicking order. \\
  \cline{3-3}
  & & Stays on message edit page after sending. \\
  \hline
  \multirow{2}{*}{Unexpected Result} & \multirow{2}{*}{App's response violates the original functional logic.} & Likes decreased by 1 after liking. \\
  \cline{3-3}
  & & Review page opened after clicking install. \\
  \hline
  \multirow{2}{*}{Data Inconsistency} & \multirow{2}{*}{Self-conflicts of data presented in the GUI.} & Ratings for the same seller are inconsistent across pages. \\
  \cline{3-3}
  & & Price does not equal original minus discount. \\
  \hline
\end{tabular}

\end{table*}

%% file: table/4.2_Interaction_perf.tex
\begin{table*}[h!]
\centering
\scriptsize
\caption{Quality comparison of converted GUI interactions.}
\label{table:interaction-perf}
\renewcommand{\arraystretch}{1}
\resizebox{1.0\linewidth}{!}{
\begin{threeparttable}
\begin{tabular}{c|cccc|cccc|cccc|cccc}
\midrule
\multirow{2}{*}{Difficulty level}    & \multicolumn{4}{c|}{L1}                                        & \multicolumn{4}{c|}{L2}                                       & \multicolumn{4}{c|}{L3}                                       & \multicolumn{4}{c}{Avg.}                                      \\
   & TC             & CS            & CT            & SE            & TC            & CS            & CT            & SE            & TC            & CS            & CT            & SE            & TC            & CS            & CT            & SE            \\ \midrule
\MIG        & \textbf{1.00} & \textbf{0.97} & \textbf{1.00} & \textbf{0.97} & \textbf{0.80} & \textbf{0.91} & \textbf{0.93} & \textbf{1.00} & \textbf{0.50} & \textbf{0.96} & \textbf{0.86} & \textbf{1.00} & \textbf{0.77} & \textbf{0.94} & \textbf{0.93} & \textbf{0.99} \\ \midrule
MobileAgent & 0.70           & 0.68          & 0.80          & 0.90          & 0.60          & 0.58          & 0.78          & 0.85          & 0.20          & 0.54          & 0.64          & 1.00          & 0.50          & 0.60          & 0.74          & 0.90          \\ \midrule
AppAgent    & 0.60           & 0.61          & 0.95          & 0.77          & 0.30          & 0.43          & 0.62          & \textbf{1.00} & 0.30          & 0.45          & 0.55          & 0.81          & 0.40          & 0.49          & 0.70          & 0.84          \\ \midrule
\end{tabular}
\end{threeparttable}
}
\end{table*}

%% file: figure/4_rq2baseline.tex
\begin{figure}
    \centering
    \includegraphics[width=0.45\textwidth]{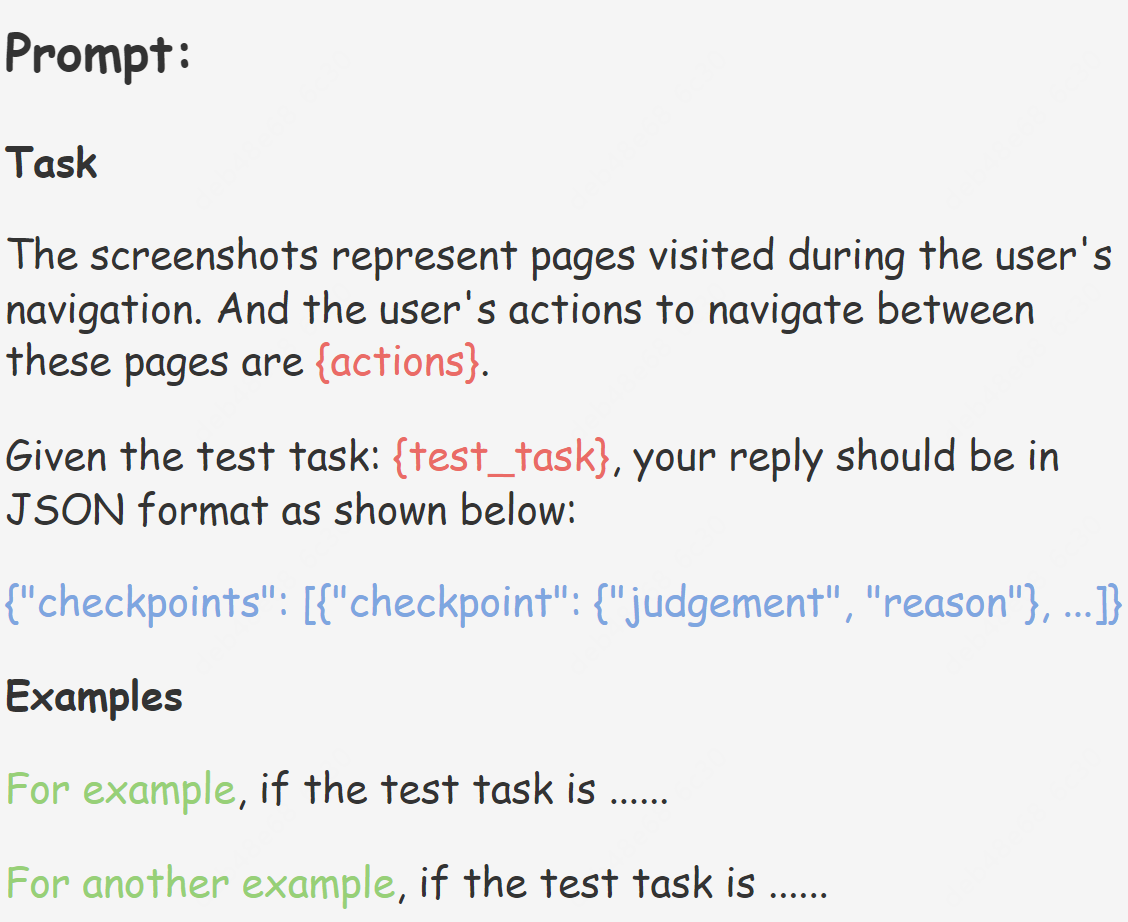}
    \caption{Prompt of RQ2 Baseline}
    \label{fig:rq2_baseline}
\end{figure}

%% file: table/4.3_veri_perf.tex
\begin{table*}[t]
\centering
\footnotesize
\caption{Effectiveness comparison of function verification.}
\label{table:veri-perf}
\renewcommand{\arraystretch}{1.2}
\resizebox{0.8\linewidth}{!}{
\begin{threeparttable}
\begin{tabular}{c|ccccc|ccccc}
\hline
\multirow{2}{*}{Task Type} & \multicolumn{5}{c|}{Correct Function Verification} & \multicolumn{5}{c}{Anomaly Detection} \\ \cline{2-11} 
  & \begin{tabular}[c]{@{}c@{}}Oracle \\ Acc.\end{tabular} & \begin{tabular}[c]{@{}c@{}}Point \\ Acc.\end{tabular} & \begin{tabular}[c]{@{}c@{}}Reason \\ Acc.\end{tabular} & \begin{tabular}[c]{@{}c@{}}Completion \\ Tokens\end{tabular} & \begin{tabular}[c]{@{}c@{}}Prompt\\ Tokens\end{tabular} & \begin{tabular}[c]{@{}c@{}}Oracle \\ Acc.\end{tabular} & \begin{tabular}[c]{@{}c@{}}Point \\ Acc.\end{tabular} & \begin{tabular}[c]{@{}c@{}}Reason\\ Acc.\end{tabular} & \begin{tabular}[c]{@{}c@{}}Completion \\ Tokens\end{tabular} & \begin{tabular}[c]{@{}c@{}}Prompt \\ Tokens\end{tabular} \\ \hline
\MIG & \textbf{0.95} & \textbf{0.97} & 0.91 & 1345 & 34987 & \textbf{0.85} & \textbf{0.91} & \textbf{0.84} & 1288 & 33715 \\
GPT-4o & 0.90 & 0.95 & \textbf{0.95} & \textbf{123} & \textbf{5695} & 0.30 & 0.53 & 0.55 & \textbf{129} & \textbf{5695} \\ \hline
\end{tabular}
\end{threeparttable}
}
\end{table*}

%% file: table/4.4_RQ3_Bug.tex
\begin{table*}[t]
\caption{RQ3-bug Information.}
\label{table:rq3-bug}
\small
\begin{tabular}{m{3cm}|m{9.6cm}|m{4cm}}
\hline
Test requirement & Bug Image & Test Result\\
\hline
Swipe up once, tap on the shop information card at the bottom of the screen (the card contains information such as the shop name and rating), then swipe down once. Check if the shop details half-screen modal appears after tapping the shop information card in step two. Verify that the shop details half-screen modal disappears after swiping down in step three. & \includegraphics[width=9.6cm]{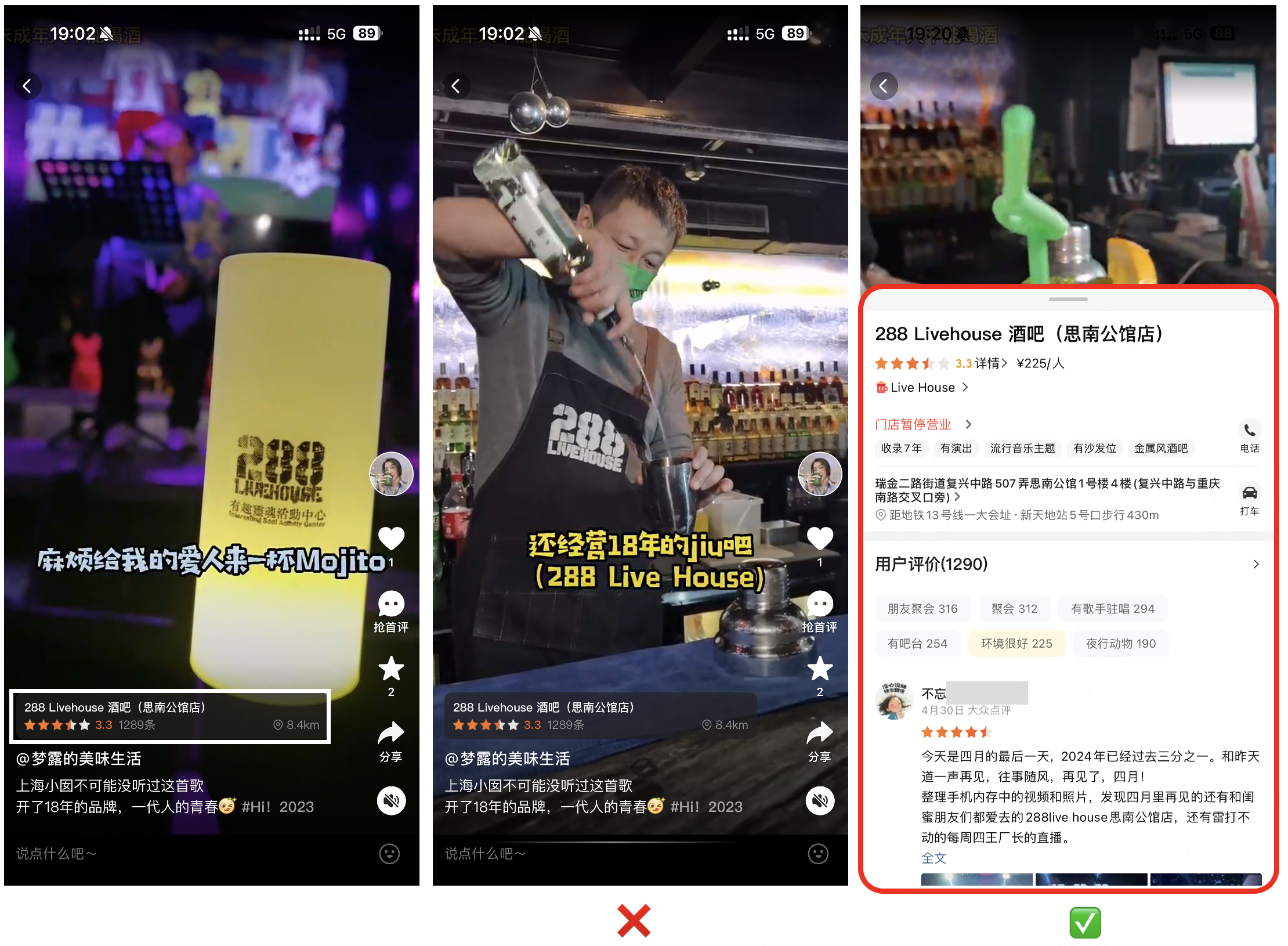}  & 
  ``reason'': ``Based on the user's navigation and interaction records, the test task was not successfully fulfilled. The records do not provide clear evidence that the store details half-popup layer was displayed or dismissed. Specifically, after the user clicked on the store information card (Action 1), there is no indication on Page 2 that the store details half-popup layer was shown. Additionally, after the user swiped down (Action 2), Page 3 does not indicate that the store details half-popup layer was dismissed. Both Page 0 and Page 3 explicitly state that the store details half-popup layer is `invisible'. Therefore, the required checks for the popup layer's visibility and dismissal were not confirmed.''
\\
\hline
\end{tabular}

\end{table*}

%% file: tex/5_Discussion.tex
\section{Threats to validity}
\label{sec:discussion}

One potential threat arises from the fact that \MIG employs a structure that decouples interaction from verification.
Specifically, in industry testing scenarios, a small number of requirements necessitate performing interactions during function verification (\eg repeatedly clicking the next video on a video playback page until a hotel-related live stream appears, then checking the details of the business to ensure the business name matches the live stream). Currently, \MIG lacks the capability to verify such requirements. 
Given their rarity, this limitation does not have much impact on \MIG’s applicability in industry settings.

Another threat stems from \MIG's reliance on general-purpose LLMs. 
General-purpose large language models such as GPT-4 are trained on extensive datasets~\cite{DBLP:journals/corr/abs-2303-08774}. However, these datasets lack domain knowledge crucial for GUI function testing, such as app function designs and GUI components.
Consequently, we acknowledge the current limitations of \MIG's capabilities. In the future, we plan to construct a dataset specific to GUI functions and enhance \MIG's proficiency in UI testing by employing a retrieval-augmented generation (RAG) approach~\cite{DBLP:conf/nips/LewisPPPKGKLYR020}.

%% file: tex/6_Related.tex
\section{Related Work}
\label{sec:related}


Since the quality of the GUI is closely related to user experience, various testing methods targeting bugs in GUI have been proposed~\cite{url:monkey,DBLP:conf/sigsoft/LamWLWZLYDX17,DBLP:conf/issta/XiongX0SWWP0023}. 
Traditionally, developing GUI testing scripts has been the main method for automating GUI function tests in industry practice. 
Since these scripts are labor-intensive to develop and are easily obsolete, several studies have been conducted on automatically generating or repairing GUI testing scripts.
For instance, AppFlow\cite{DBLP:conf/sigsoft/HuZY18} utilizes machine learning techniques to automatically identify screen components, enabling testers to write modular testing libraries for essential functions of applications. CraftDroid\cite{DBLP:conf/kbse/LinJM19} uses information retrieval to extract human knowledge from existing test components of applications to test other programs. CoSer\cite{DBLP:conf/icse/CaoP000WL24} constructs UI state transition graphs by analyzing app source code and test scripts to repair obsolete scripts.
Although \MIG is also a method for testing GUI functions, it differs from these studies as it neither relies on nor generates hard-coded testing scripts, thereby offering better generalizability and maintainability.

Due to their extensive scale of training data and robust logical reasoning abilities, LLMs are increasingly utilized in mobile app testing.
Several methods are proposed using LLMs to generate GUI interactions or translating natural language commands into GUI actions.
QTypist~\cite{DBLP:conf/icse/LiuCWCHHW23}  focuses on generating semantic textual inputs for form pages to enhance exploration testing coverage.
GPTDroid~\cite{DBLP:conf/icse/0025C0CWCW024} extracts GUI page information and widget functionality from the UI hierarchy file, using this data to generate human-like interactions.
AppAgent~\cite{DBLP:journals/corr/abs-2312-13771}, Mobile-Agent~\cite{DBLP:journals/corr/abs-2401-16158}, DroidBot-GPT~\cite{DBLP:journals/corr/abs-2304-07061} and AutoDroid~\cite{DBLP:journals/corr/abs-2308-15272} leverage LLMs to process natural language descriptions and GUI pages, translating natural language commands into GUI actions.
Different from these tools, \MIG can directly perform GUI testing on mobile apps. Moreover, our experiments demonstrate that the techniques employed by \MIG, including dynamically organizing agents, lead to the highest quality of GUI interaction generation.

VisionDroid~\cite{liu2024visiondrivenautomatedmobilegui} focuses on non-crash bug detection of GUI pages, highlighting the absence of testing oracles and utilizing large language models (LLMs) to detect unexpected behaviors. 
In contrast, \MIG concentrates on the industry's practical needs. By emphasizing the challenges of implementing practical testing requirements, we have developed an industry-applicable automatic natural language-driven GUI testing method.